\documentclass[twocolumn,secnumarabic,amssymb, aps, prl, superscriptaddress]{revtex4-1}
\usepackage{graphicx}
\usepackage{dcolumn}
\usepackage{bm}
\usepackage{xcolor}
\usepackage{lipsum}
\usepackage{upgreek}
\usepackage{float}
\usepackage{graphicx}
\usepackage{amsmath}
\usepackage[colorlinks, citecolor=blue]{hyperref}

\begin{document}

\title{Quadrature Photonic Spatial Ising Machine}

\author{Wenchen Sun}
\affiliation{State Key Laboratory of Advanced Optical Communication Systems and Networks,Shanghai Jiao Tong University, Shanghai, China 200240}

\author{Wenjia Zhang}
\email{wenjia.zhang@sjtu.edu.cn}
\affiliation{State Key Laboratory of Advanced Optical Communication Systems and Networks,Shanghai Jiao Tong University, Shanghai, China 200240}

\author{Yuanyuan Liu}
\affiliation{State Key Laboratory of Advanced Optical Communication Systems and Networks,Shanghai Jiao Tong University, Shanghai, China 200240}

\author{Qingwen Liu}
\affiliation{State Key Laboratory of Advanced Optical Communication Systems and Networks,Shanghai Jiao Tong University, Shanghai, China 200240}

\author{Zuyuan He}
\affiliation{State Key Laboratory of Advanced Optical Communication Systems and Networks,Shanghai Jiao Tong University, Shanghai, China 200240}

\begin{abstract}
The mining in physics and biology for accelerating the hardcore algorithm to solve non-deterministic polynomial (NP) hard problems has inspired a great amount of special-purpose machine models.
Ising machine has become an efficient solver for various combinatorial optimization problems.
As a computing accelerator,
large-scale photonic spatial Ising machine have great advantages and potentials due to excellent scalability and compact system.
However, current fundamental limitation of photonic spatial Ising machine is the configuration flexibility of problem implementation in the accelerator model.
Arbitrary spin interactions is highly desired for solving various NP hard problems.
Moreover, the absence of external magnetic field in the proposed photonic Ising machine will further narrow the freedom to map the optimization applications.
In this paper, we propose a novel quadrature photonic spatial Ising machine to break through the limitation of photonic Ising accelerator by synchronous phase manipulation in two and three  sections. 
Max-cut problem solution with graph order of 100 and density from 0.5 to 1 is experimentally demonstrated after almost 100 iterations. 
We derive and verify using simulation the solution for Max-cut problem with more than 1600 nodes and the system tolerance for light misalignment.
Moreover, vertex cover problem, modeled as an Ising model with external magnetic field, has been successfully implemented to achieve the optimal solution. 
Our work suggests flexible problem solution by large-scale photonic spatial Ising machine.

\end{abstract}

\maketitle

\begin{figure*}[t!]
\centering  
\includegraphics[width=17cm]{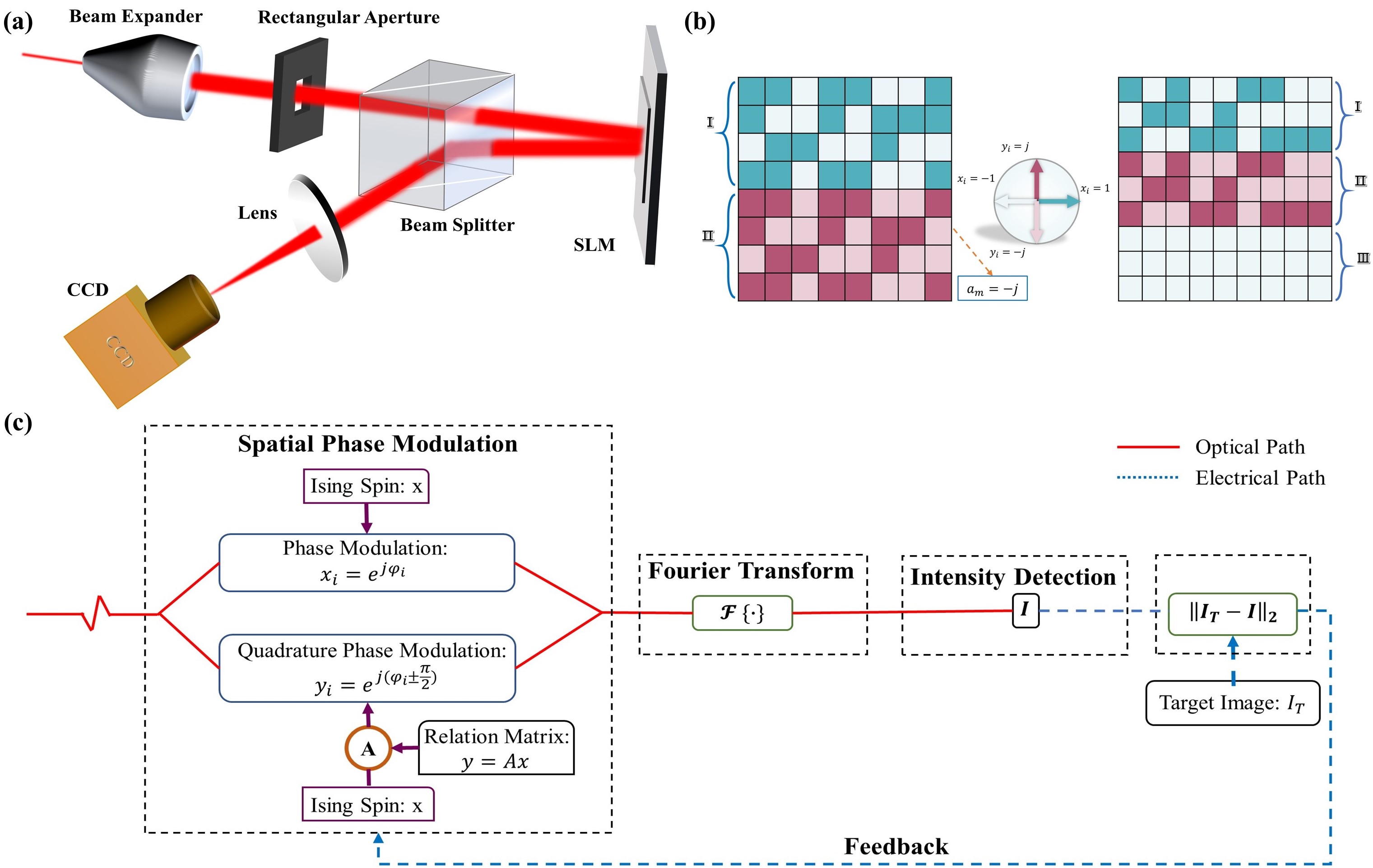} 
\caption{
\textbf{Architecture of quadrature photonic spatial Ising machine.} (a) Experimental setup. Spatial phase of an expanded laser beam with rectangular spot is modulated by  $\{0,\pi\}$ or $\{\frac{\pi}{2}, \frac{3\pi}{2}\}$ at two places. 
Intensity profile of phase-modulated light shining on an image sensor through two-dimensional Fourier transform performs the Hamiltonian of Ising model.
(b) The phase maps loaded to SLM. 
The left phase map shows two-section phase configuration with orthogonal relation, 
while the right one presents a three-section phase configuration to characterize a Hamiltonian with external magnetic field.
(c) The flow chart of optimization process. 
} 
\label{CouplingPrinciple}
\end{figure*}

\section{Introduction}
The innovative application-specific hardware in the platforms of quantum mechanics \cite{jiuzhang}, memristors \cite{tianji} and photonics \cite{Shastri2018,DDNN,yichen-NP} are attracting enormous attentions in order to solve large-scale computation-intensive problems that cannot work out timely and efficiently by conventional electronic architectures in the post-Moore era. 
Combinatorial optimization, most of which are classified as non-deterministic polynomial (NP) hard, is one of important but computational intractable problems and find their critical applications in artificial intelligence, scheduling, finance prediction and $etc$ \cite{bookcomb,lucas:14}.
However, typical NP hard problems cannot be tackled in the conventional computing architecture with a reasonable cost as training data and adjustable parameters grow exponentially for a real-world challenge.
The analogy between the phenomenon in physics and biology and computer science has inspired a great amount of special-purpose machine models that are used to accelerate the hardcore algorithm in figuring out an available result.
Ising machine, originated from Ising model describing a lattice of sites ${l}$ with a single but two-state degree of freedom $x_l$ called a spin on each site by taking values of $\pm{1}$, is a dynamics physical system that evolve towards the minimum Hamiltonian in \eqref{Ising Hamiltonian} by continuous change of spin binary state $x_{l} \in \{1,-1 \}$. 
\begin{equation}
    H=-\sum_{<l,k>}J_{l,k}x_{l}x_{k}-\sum_{k}h_{k}x_{k}
    \label{Ising Hamiltonian}
\end{equation}
where $J_{l,k}$ is the interaction between spins and $h_{k}$ is spin interact with external magnetic field. 
The property of the one and two-dimensional Ising model without magnetic field have been analytically solved by Ising and Onsager in a complicated way \cite{Isingone,Onsagertwo} . 
With the emergence of high-performance computing system, the Monte Carlo method, by forming a Markov chain as a system evolves in steps, is employed by doing random spin configuration in order to minimize the Hamiltonian. 
Therefore, the map of Ising machine with various optimization problems has been naturally conceived so that the engineering procedure with a large amount of parameters can be managed by an equivalent Hamiltonian through flipping the state of a spin.  
Owning to the advantage of inherent parallelism, scalability, and power efficiency, optical computing architecture rekindles the spark in academia by proposing various solutions for emerging artificial intelligence and is believed to play an indispensable role on innovative computing platform for future specific applications \cite{Shastri2018,DDNN,yichen-NP, Alexoudi:19}.
Through remarkable innovations,  NP hard problems including subset sum problem \cite{Xueaay5853}, dominating set problem \cite{Goliaei:12} and Hamiltonian path problem \cite{Vazquez:18}, have been tackled efficiently by photonic computing.
Photonic Ising machines, with spin state encoded to phase or intensity of light, have been proposed by leveraging optical parametric oscillators \cite{PhysRevA.88.063853,marandi2014network,Mcmahon:16,Inagaki603}, integrated linear photonics \cite{MIT}, and spatial field manipulation \cite{PRL}. 
These machines show similar computational acceleration as previous demonstrations \cite{Shastri2018,DDNN,yichen-NP} but have the profound difference from the traditional matrix-vector multiplication where Hamilton is calculated by \eqref{Ising Hamiltonian}.
Networks of degenerate optical parametric oscillators (DOPOs) are employed as a physical platform with an unconventional operating mechanism to construct a coherent Ising machine \cite{PhysRevA.88.063853}. 
Complicated controlling circuit implemented in field-programmable gate array (FPGA) enables successful manipulation a coupling matrix with 100 spins \cite{haribara2015coherent}.
Recently, short microwave pulses from DOPO is utilized to increase the system scalability and computing accuracy \cite{MPising}.
DOPOs is the first solution to realize a large-scale Ising machine but bulky structure with discrete devices requires complex software for stabilizing phase transition and coupling. 
Therefore, a coherent integrated network consisted of Mach-Zehnder interferometers (MZIs) has been utilized to achieve an optimum solution for an arbitrary Ising model powered by  an iterative heuristic algorithm \cite{MIT}.
With precise control, Max-cut problem with various graph densities and 100 graph order can be solved within a millisecond.
However, although this proposal, intrinsically behaving as an accelerator of vector-matrix multiplication, provides flexibility for problem configuration,  it is hindered by large number of iterations and integration scalability due to imperfect components \cite{MIT,Miller:15}.
Diffractive component has shown very unique feature due to spatial multiplexing and will present profound influence on large-scale optical computing \cite{PRL,Lin1004,luo2019design,zhou2020large,chang2018hybrid,zhou2020situ}. 
A compact optical setup can carry out various tasks, such as image classification  \cite{chang2018hybrid} and optical neural network construction \cite{zhou2020situ}. 
Spatial light modulator (SLM) as a reconfigurable diffractive device with phase control in micrometer resolution is widely adopted for light field manipulation and thereby to implement a large-scale Ising machine \cite{PRL,Pierangeli:20}.
The phase of a pixel in the SLM characterize as a spin and interaction matrix is the product of intensity through Fourier optical imaging system.
Though the scale of Ising model can be upgraded to tens of thousands \cite{PRL}, the very rigid structure through spatial light interference limits flexible configuration for a real-world problem. 
However, arbitrary interaction configuration is highly desired for solving NP hard problem. 
Moreover, current Ising machines are all absent of external magnetic field in Ising model, which will simplify the minimized Hamiltonian searching algorithm but narrow the freedom to map various optimization applications.

In this paper, we propose a quadrature photonic spatial Ising machine (Q-SIM) by synchronous phase manipulation in two and three  sections.
Arbitrary negative spin interaction can be configured through tuning the light amplitude and spatial phase based relation matrix in the proposed architecture.
Max-cut problem solution with graph order of 100 and density from 0.5 to 1 is experimentally demonstrated after almost 100 iterations.
We also derive and verify using probabilistic simulation the solution for Max-cut problem  with  more  than  1600  nodes  and  the system tolerance for light misalignment. 
Moreover, vertex cover problem, modeled as a Ising model with external magnetic field, has been successfully implemented to achieve the optimal solution, which for best our knowledge cannot be solved by any photonic Ising machine. 

\section{methods}
Architecture of quadrature photonic spatial Ising machine is shown in Fig. \ref{CouplingPrinciple}(a). A beam of light intensity before spatial phase modulator can be expressed as 
\begin{equation}
   P(u,v)=\sum_{l}\sum_{k}\xi_{(u_{l},v_{k})}rect(\frac{u-u_{l}}{W})rect(\frac{v-v_{k}}{W})
\end{equation}
where $\xi_{(u_{l},v_{k})}$ is the intensity of a pixel, $(u_{l},v_{k})$ is a central location of pixel, the size of pixel is $W\times W$. 

After spatial phase modulation, this beam becomes
\begin{equation}
     Q(u,v)=\sum_{l}\sum_{k}\xi_{(u_{l},v_{k})}x_{(u_{l},v_{k})}rect(\frac{u-u_{l}}{W})rect(\frac{v-v_{k}}{W})
\end{equation}
where $x_{(u_{l},v_{k})}=e^{k\varphi_{(u_{l},v_{k})}}\in \{-1,1\}$ corresponds to the value of modulated phase $\{0,\pi \}$ in the pixel $(u_{l},v_{k})$.

Finally, central intensity from spatial intensity distribution can be obtained at the charge-coupled device (CCD) image sensor after two-dimensional Fourier transform of the emergent light.

\begin{equation}
    I(0,0)=W^{4}{\sum_{r,l}^{r\ne l}\sum_{s,k}^{s\ne k}\xi_{(u_{l},v_{k})}\xi_{(u_{r},v_{s})}x_{(u_{l},v_{k})}x_{(u_{r},v_{s})}+constant}
    \label{intensity}
\end{equation}
To facilitate analysis, we can rewrite \eqref{intensity} as
\begin{equation}
    I(0,0)=x^{T}\xi\xi^{T}x+constant
    \label{vema}
\end{equation}
where $x^{N}=[x_{(u_{1},v_{1})},x_{(u_{1},v_{2})},...,x_{(u_{M},v_{M})}]\in\{-1,1\}^{N}$ and $\xi^{N}=[\xi{(u_{1},v_{1})}, \xi{(u_{1},v_{2})},...,\xi{(u_{M},v_{M})}]$. 

Therefore, it is inversely proportional to the Hamiltonian of an Ising model with interaction matrix of $J^{N\times N}=\xi\xi^{T}$, which is the foundation of spatial Ising model \cite{PRL}.
However, non-negative $\xi$ limits Ising model in the ferromagnetic form.
Moreover, it is infeasible to construct a NP hard problem with incomplete graph that some interactions are forced to zero, leaving others in the same node intact. 
For instance, if $\xi{(u_{l},v_{k})}=0$, all $J_{l,k}$ calculated through multiplying $\xi{(u_{l},v_{k})}$ will equal to zero where $J_{l,k}$ is the interaction between $l-th$ spin and $k-th$ spin.
This is the fundamental limitation of photonic spatial Ising model that hinders its practical implementation in a NP hard problem solver.  
Although the choice of target image \cite{PRL}, to some extent, can break this limitation, there still exists an inherent problem.  
For instance, by introducing a target image with Fourier transformation of $\widetilde{I}_{T}$, $J_{l,k}$, according to \cite{PRL}, can be configured by
\begin{equation}
    J_{l,k}=2\pi\xi_{l}\xi_{k}\widetilde{I}_{T}(2W(l-k))
\end{equation}
In this case, if $J_{1,2}$ is required to be zero, thereby arbitrary $J_{l,l+1}$ will be forced to zero because $\widetilde{I}_{T}(-2W)$ is zero.
To improve configuration flexibility, we propose a novel quadrature photonic spatial Ising machine, featured with spatial coded phase modulation. 
As shown in Fig. \ref{CouplingPrinciple}(c), two Ising models with designed relation are constructed and updated in a synchronous state. 
The relation matrix $A=diag(a_{1}, a_{2},...,a_{N})$, determining the spin value of second Ising model, is a unitary diagonal matrix and satisfy
\begin{equation}
    a_{l}=e^{j\theta_{l}}
\end{equation}
As $\theta_{l}$ is regarded as modulated phase added to the second Ising model, spins will rotate $\theta_{l}$, which can be called by $\theta$-spin.
The original spin and $\theta$-spin value vector of the first and second Ising model are denoted by $x$ and $y$ respectively, which satisfies
\begin{equation}
    y=Ax
    \label{phaseRelation}
\end{equation}
In the image sensor, the center intensity can be detected as
\begin{equation}
    I(0,0)=(x^{T}\xi+y^{T}\eta)(\overline{\xi^{T} x+\eta^{T}y} )
    \label{twoIntensity}
\end{equation}
where $\xi$ and $\eta$ are spatial intensity distribution respectively.

\begin{figure}
    \centering
    \includegraphics[width=8cm]{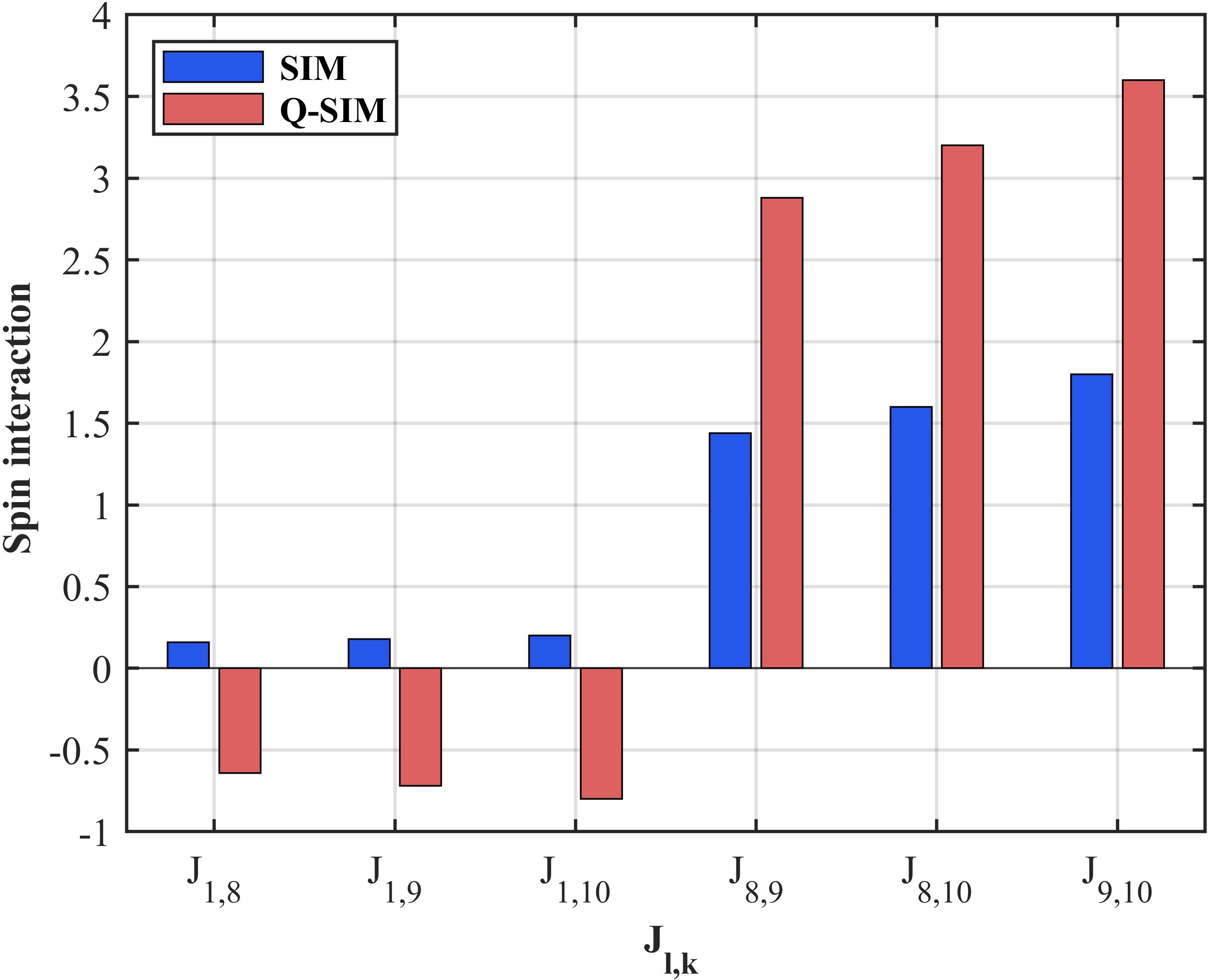}
    \caption{\textbf{An example of spin interaction between quadrature spatial photonic Ising machine (Q-SIM) and original spatial photonic Ising machine (SIM).} 
    By introducing the relation matrix $A$ that composed of half $j$ and $-j$,
    the interaction values are able to change from non-negative to arbitrary values through tuning the light amplitude and relation matrix.
    }
    \label{fig:interactionJ}
\end{figure}

By utilizing \eqref{phaseRelation}, \eqref{twoIntensity} represents an Ising model with 
\begin{equation}
    J_{l,k}=2\xi_{l}\xi_{k}+2Re\{a_{l}\}\xi_{k}\eta_{l}+2Re\{a_{k}\}\xi_{l}\eta_{k}+2Re\{a_{l}\overline{a_{k}}\}\eta_{l}\eta_{k}
    \label{constructing J}
\end{equation}
where $Re\{\cdot\}$ is used to take the real part.
If $A=diag(j,..,-j,...,j)$ where the index of $-j$ is r, the $\theta$-spin will become a quadrature-spin and 
\begin{equation}
    J_{l,k}=
    \begin{cases}
    2\xi_{l}\xi_{k}-2\eta_{l}\eta_{k} &, for\ l=r\ or\ k=r\ and\ l \neq k\\
    2\xi_{l}\xi_{k}+2\eta_{l}\eta_{k} &, others\\
    \end{cases}
    \label{oneZeros}
\end{equation}
An extra dimension helps us to configure interaction in a larger range compared to the previous architecture. 
For instance, in order to build an incomplete graph or anti-ferromagnetic spin glass system, we can tune the intensity $\xi$ and $\eta$ of \eqref{oneZeros} to get $\xi_{r}\xi_{k}=\eta_{r}\eta_{k}$ or $\xi_{r}\xi_{k}<\eta_{r}\eta_{k}$.
Fig. \ref{fig:interactionJ} shows spin interactions of Q-SIM where we set the relation matrix as $diag(-j,-j,...,j)$ with equal number of  $j$ and $-j$ and same intensity configuration to $\xi$ and $\eta$.
As we can see from this figure, the values of interaction $J_{l,k}$ are able to change from non-negative of original spatial Ising machine to arbitrary value through tuning the light amplitude and relation matrix in the proposed architecture.

The Q-SIM can be further extended to three spatial pixel blocks in order to embrace the case with external magnetic field as shown in \eqref{Ising Hamiltonian}. 
Spatial phase with values of $\{0,\pi\}$ and $\{\frac{\pi}{2},\frac{3\pi}{2}\}$ is modulated in the first two parts, leaving last part a fixed phase.
Therefore, the central intensity from CCD detector can be expressed as 
\begin{equation}
   I(0,0)=x^{T}Jx+2z^{T}\sigma\xi^{T}x 
   \label{eq:splitThreeParts}
\end{equation}
where $x$ and $z$ are the first and third Ising spin value with $\xi$, $\eta$ and $\sigma$ as different modulated intensities and $J$ as \eqref{constructing J}. 
The extension including external magnetic field will contain various NP hard problems, such as vertex cover problem and traveling salesman problem.
For solving those NP hard problems, we firstly tune $\xi_{k}$  to satisfy the $h_{k}$ value. With fixed $\xi_{k}$, $\eta_{l}$ can be tuned to achieve different values for the required matrix.


\section{Experiment and result}
Experimental setup is shown in Fig. \ref{CouplingPrinciple}(a). 
Intensity modulated incident beam with $\lambda=632.8  nm$ and output power of $1.6 mW$ shines on a reflective SLM of HOLOEYE PLUTO-NIR-011 with $1920\times1080$ pixels. The image frame rate is 60Hz and a single pixel size is $8 \mu m\times 8 \mu m$. 
Thereby, spatial phase of an expanded laser beam with rectangular spot is modulated by  $\{0,\pi\}$ or $\{\frac{\pi}{2}, \frac{3\pi}{2}\}$. 
After spatial phase modulation, intensity profile of this beam is transformed through two-dimensional Fourier optics with focal length of $15cm$ and detected on CCD camera with frame rate of $40Hz$ and quantization bit of 8. 
For controlling feedback to search ground state of arbitrary Ising model, we randomly flip every spin to approach a target image with intensity decreasing stepwisely from the center to the outside.
Due to the slow speed of SLM and CCD, the iteration time for the experiment is set to $1s$ with five times intensity detection in order to obtain an averaging intensity output.
Specifically, for each iteration, difference between detected intensity image $I$ and target image $I_{T}$ is calculated as
\begin{equation}
    d=||I-I_{T}||
\end{equation}
After flipping the spatial phase, the difference of detected new intensity image $I_{new}$ and target image $I_{T}$ is calculated as 
\begin{equation}
    d_{new}=||I_{new}-I_{T}||
\end{equation}
For minimizing the Hamiltonian of Ising model, we will keep changing spatial phase only until $d_{new}<d$ so that the detected image is gradually approaching the target image.

\subsection{Fully connected Ising model}

We introduce negative interaction to observe the physical property of spin glass system.
Magnetization, defined by $\sum_{l=1}^{N} x_{l}/N$, is used to measure the level of randomness for a physical system. 
Thanks to the adoption of synchronous quadrature Ising model,  negative interaction can be configured according to the concrete applications. 
By \eqref{constructing J}, if relation matrix is composed of $j$ and $-j$ 
and intensity $\eta$ is greater than intensity $\xi$, the sign of $J_{lk}$ will be decided by $a_{l}\overline{a_{k}}\eta_{l}\eta_{k}$. 
Suppose the number of $-j$ is $r$, the number of negative interaction will be $NI=r(N-r)$ where $N$ is the spin number. Negative interaction, causing spin state opposite, can be achieved by the product of $-j$ and $j$. 
The magnetization is
\begin{equation}
    m=\pm\frac{N-2r}{N}
\end{equation}
When we take absolute value of $m$, the relation between $NI$ and $|m|$ can be derived
\begin{equation}
    |m|=\sqrt{1-\frac{4NI}{N^{2}}}
    \label{Eq:|m|}
\end{equation}

\begin{figure}
    \centering
    \includegraphics[width=\columnwidth]{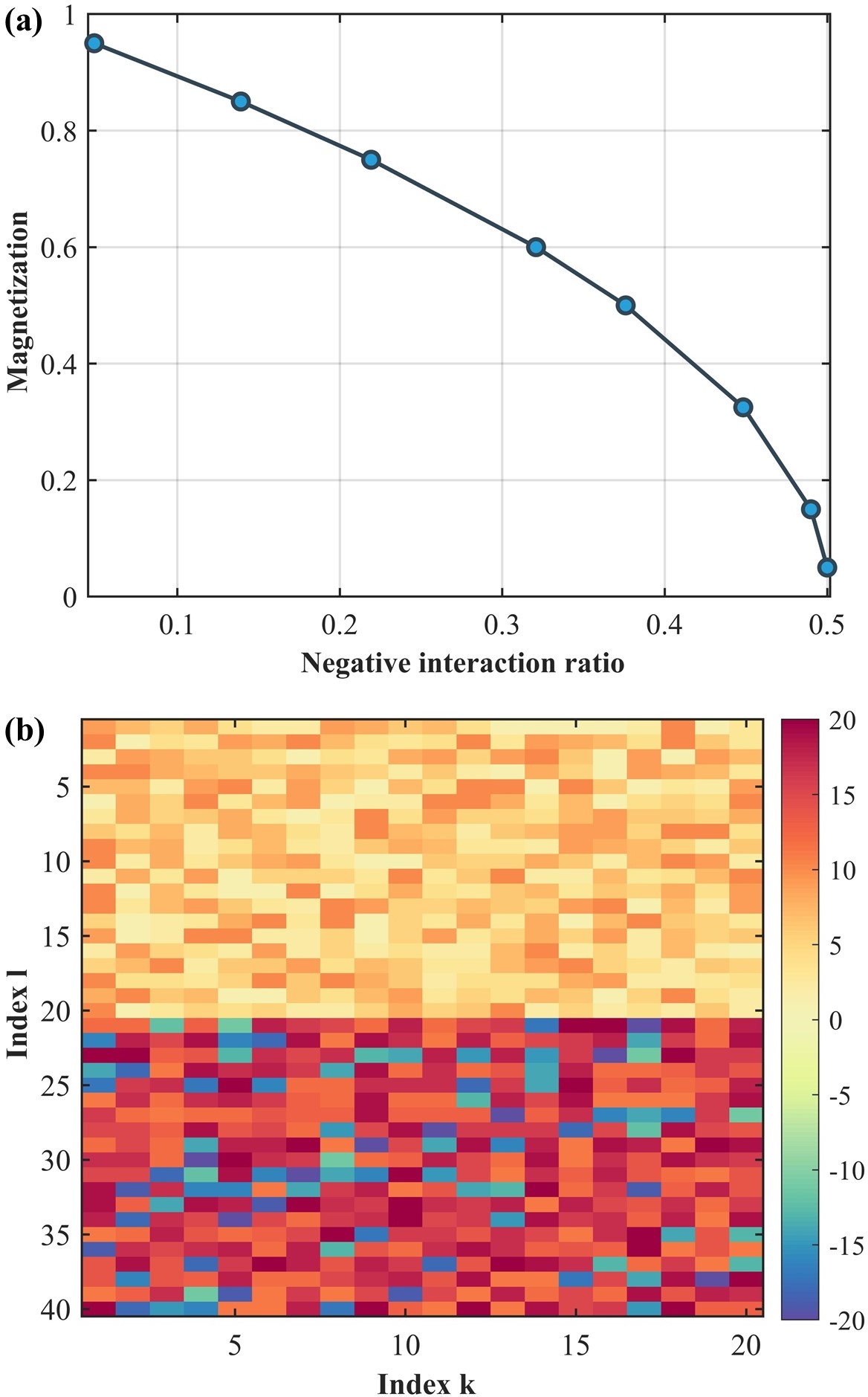}
    \caption{\textbf{The physical property of spin glass system.} (a) Magnetization of ground state versus negative interaction ratio. 
    (b)
    Spatial intensity and relation matrix of the beam when negative interaction ratio is 0.3208. In the second section, negative and positive value corresponds to -j and j respectively in relation matrix.}
    \label{fig:firstPart}
\end{figure}

For a model with 400 spins, Q-SIM model with negative interaction is evaluated through numerical simulation in order to confirm its effect to final state of spins.
The number of negative interaction is normalized to the ratio of negative interaction to total interaction in the Ising model, which  can be tuned by updating relation matrix in the Q-SIM model.
As shown in Fig. \ref{fig:firstPart}(a), magnetization monotonically decrease with negative interaction ratio, which well agrees with \eqref{Eq:|m|}.
The negative interaction expands solvable problem space and will be significant for various problem configurations. 
To facilitate this process, a mapping algorithm, from NP hard problem to Q-SIM, is required to obtain loaded intensity profile of incident light and relation matrix.
For instance, when negative interaction ratio is 0.3208, the required spatial intensity of a beam can be figured out and the number of $-j$ in relation matrix is 80 as shown in the Fig.\ref{fig:firstPart}(b).
\subsection{Solving Max-cut problem}
Max-cut problem of finding a maximum cut in a graph can be mapped to Ising model\cite{garey1979computers}. 
In the Q-SIM, graph node is represented by a spin consisting of pixel block and the modulated phase indicates which vertices subset that this node belong to.
Edge weight can be configured by intensity profile denoted by $\xi$ and $\eta$ in every pixel.
\begin{figure}[t]
\centering  
\includegraphics[width=\columnwidth]{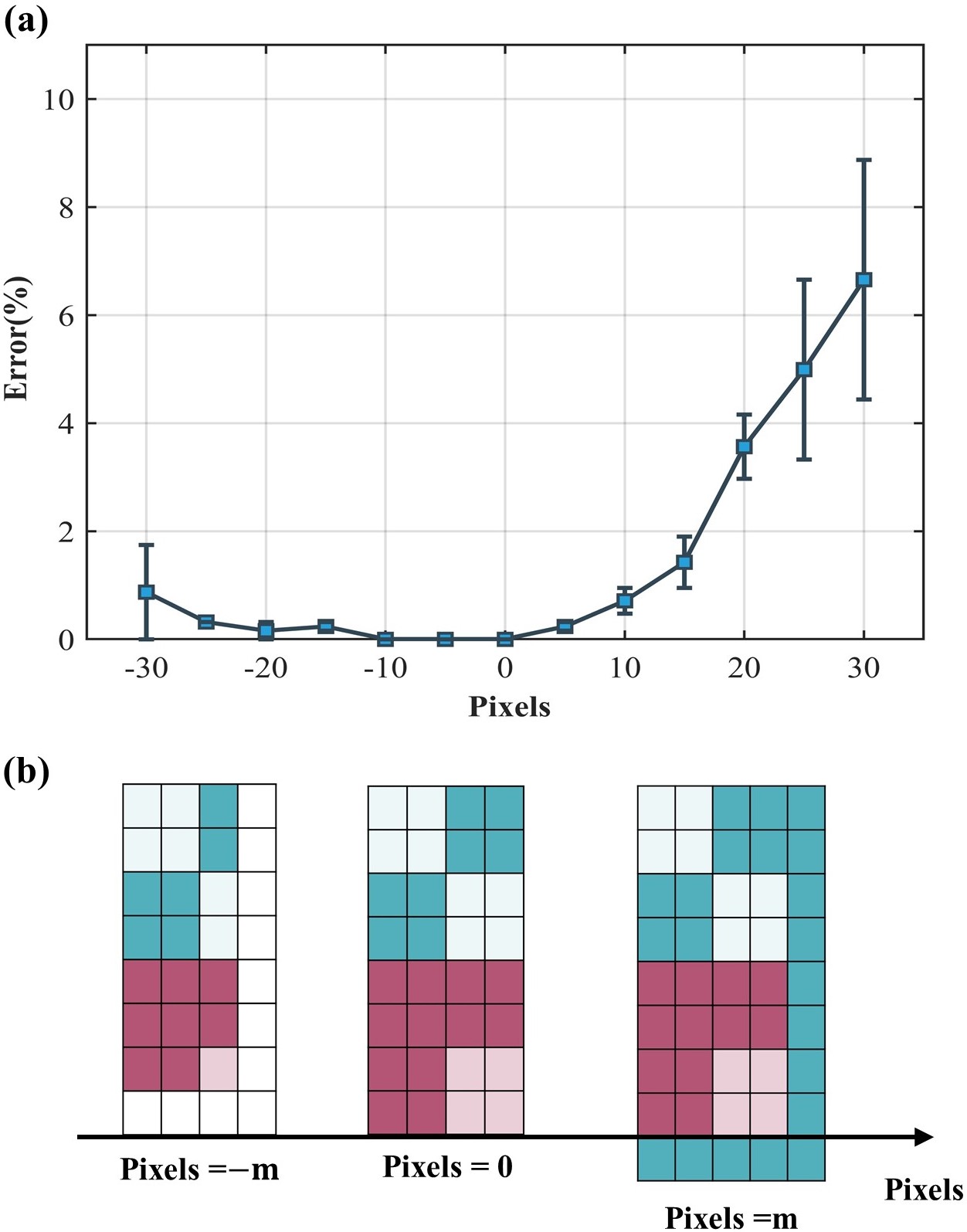} 
\caption{\textbf{ Performance evaluation with light misalignment.} (a) Error deviation of cut value after 76 iterations. (b) Illustration of light misalignment for quadrature modulation in two areas.
}
\label{thethirdpart}
\end{figure}
\begin{figure*}[t!]
\centering  
\includegraphics[width=1.95\columnwidth]{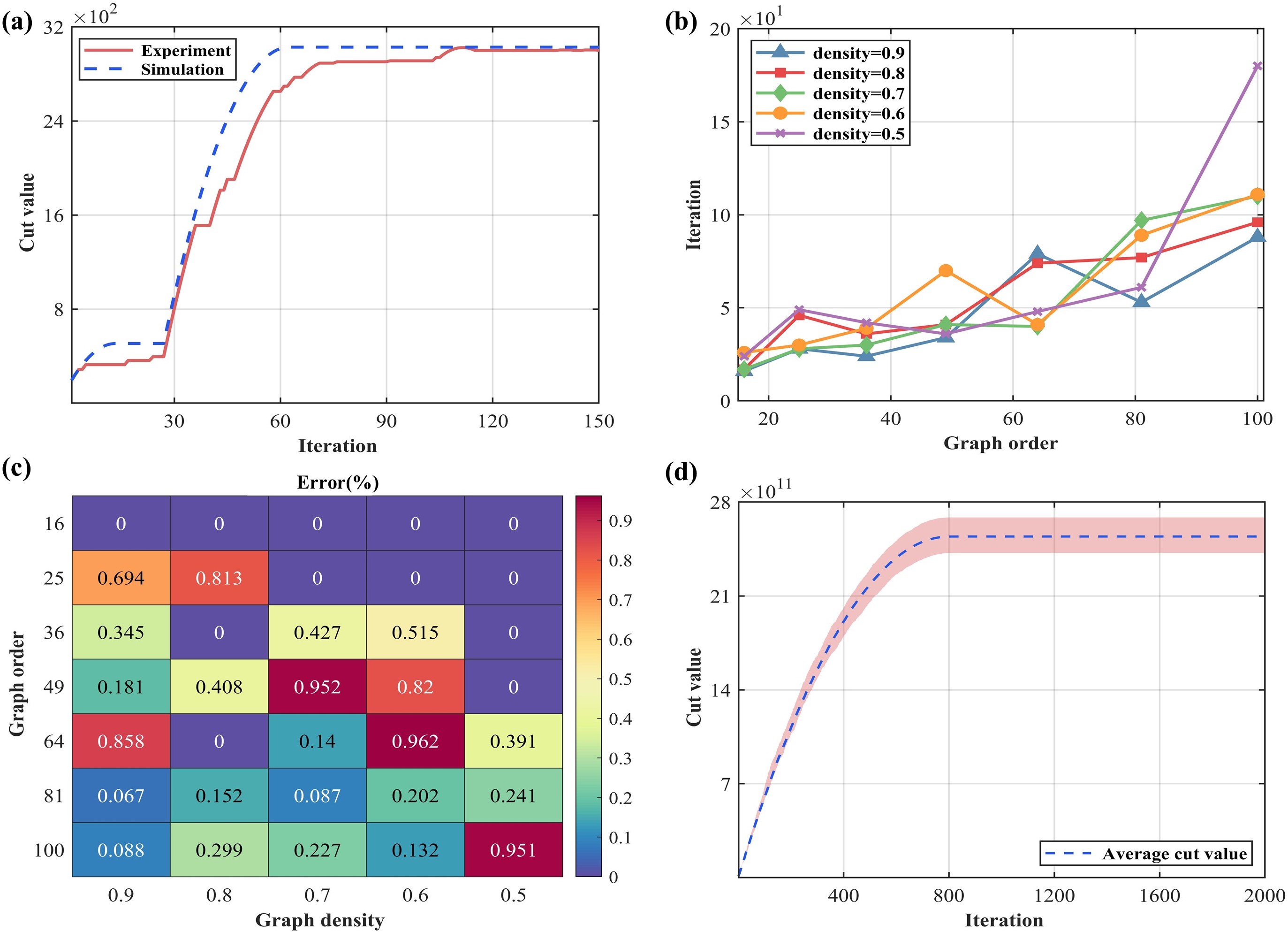}
\caption{\textbf{Solving Max-cut problem in arbitrary weight.} (a) The cut value of a 100-order graph with graph density of 0.6. 
(b) The required iteration number when obtaining the Max-cut solution with various graph density and order. 
(c) The solution deviation compared to optimum. 
(d) Confidence interval of cut value by randomly generating 50 weighted 1600-nodes graphs. 
}
\label{theSecondpart_experiment}
\end{figure*}
With proper graph configuration, Max-cut problem is equivalent to find ground state of Ising model. 
Therefore, the intensity at the center of CCD is inversely proportional to weight sum by \eqref{twoIntensity}. 
Iterative algorithms based on binary search is utilized to away the intensity profile to a target image. 
The result of a Max-cut problem will be achieved through reading the finalized phase distribution in SLM. 

Max-cut problem with incomplete graph is experimentally solved through Q-SIM implementation. 
In order to construct incomplete graph, phase of $\{0,\pi\}$ is modulated on top of first section in SLM and $\{\frac{\pi}{2},\frac{3\pi}{2}\}$ for second section.
Moreover, relation matrix needs to be constructed for realizing various graph densities.
Take a 100-order graph with graph density of 0.6 as an example, relation matrix can be set $A=diag(-j,-j,...,j)$ where the number of $-j$ and $j$ is 27 and 73 respectively. 
Fig. \ref{theSecondpart_experiment}(a) shows cut value increases with iteration number, converging to a stable value both by simulation and experiment. 
The experimental results show the validity and efficiency of Q-SIM to solve Max-cut problem.
With the merit of spatial diversity, iteration required for achieving maximum cut value, about 100, is much less than integrated photonic Ising machine in the same scale \cite{MIT}.    
We also design graphs with graph density of $0.6\sim0.9$, graph order of $16\sim100$ and edge weight of $1$ or $0$.
Fig. \ref{theSecondpart_experiment}(b) illustrates the convergence iteration number for each case with $99\%$ confidence of optimum partition.
From Fig. \ref{theSecondpart_experiment}(b), iteration number in the graph order of 100 is approximately similar for different graph densities.
We show the achieved max cut compared to the optimized one 
for Max-cut problem with different graph orders and densities in Fig. \ref{theSecondpart_experiment}(c). As Fig. \ref{theSecondpart_experiment}(c) shows, most max cut deviation is lower than 1\% in the experiment.
In order to improve the universality of this approach, we randomly generate a 1600-order graph with weights ranging from $-4\times10^{6}$ to $2\times10^{7}$. 
Fig. \ref{theSecondpart_experiment}(d) shows cut values are swiftly increasing with iterations for different random generated cases and tending to flat after 800 iterations, proving that Max-cut problem can be efficiently solved in the scale of at least 1600-order graph. 
Moreover, Fig. \ref{theSecondpart_experiment}(d) also indicates the similar computing performance can be achieved for graphs in the same graph order but with different weights.

The system noises, originated by intensity discretization and processing in the CCD and imperfect spatial light modulation, have been comprehensively analyzed in \cite{PierangeliMarcucciBrunnerConti+2020+4109+4116} and will help to avoid the system trapping in local energy minima. 
With fixed phase deviation and exposure time of CCD, we focus on the issue of optical misalignment in the vertical direction of the SLM plane, which is critical for configuring the quadrature modulation.
When light beam is expanded and shaped through geometrical optics with rectangular aperture, this aperture needs strictly same as the spot size for phase modulation. 
As shown in Fig. \ref{thethirdpart}(b), if the vertical distance from rectangular aperture to SLM is not properly controlled, the light field size will expand larger or shrink smaller than the size of effective SLM for quadrature modulation.   
$Pixels<0$ means phase pixels located at last m columns and m rows are inactive due to small light spot while $Pixels>0$ means opposite.
Therefore, we create using simulation a Max-cut problem with graph order of 100 and graph density of $0.5$ for evaluating this misalignment issue. 
The spin node is characterized by $30\times 30$ pixels with the pixel size of $8\mu m$ and the change of size error is controlled within  30 pixels.
Note that performance error is defined by cut value deviation after 76 iterations so that the cut value will approach to optimum value for the best case. 
As shown in Fig. \ref{thethirdpart}(a), both the average and error deviation statically increases with the light misalignment and the computational error is less than 2\% when the light field is decreased by 30 pixels, which is better than the case with light field expansion since that expansion will cover more phase pixels without optimization.

\subsection{Vertex cover problem}
Vertex cover problem of finding a vertex subset to contain at least one vertex of edge set is a NP hard problem.
In vertex cover problem, vertex is corresponding to spin and binary spin value determines whether vertex is in the subset \cite{lucas:14}.
Then solving vertex cover problem can be mapped to find ground state of Ising model with external magnetic field.
Therefore, we need a three-section phase configuration to characterize a Hamiltonian with external magnetic field.
By \eqref{eq:splitThreeParts}, we can set
\begin{equation}
    A=2(\xi_{l}\xi_{k}\pm\eta_{l}\eta_{k})
\end{equation}
\begin{equation}
    B-(N-1)A=2\xi_{l}\sum_{l}z_{l}\sigma_{l}
\end{equation}
When spatial intensity ($\xi_{l}$, $\eta_{l}$, $\sigma_{l}$) are properly configured, vertex cover problem can be solved by our Q-SIM. 

\begin{figure}
\centering
\includegraphics[width=\columnwidth]{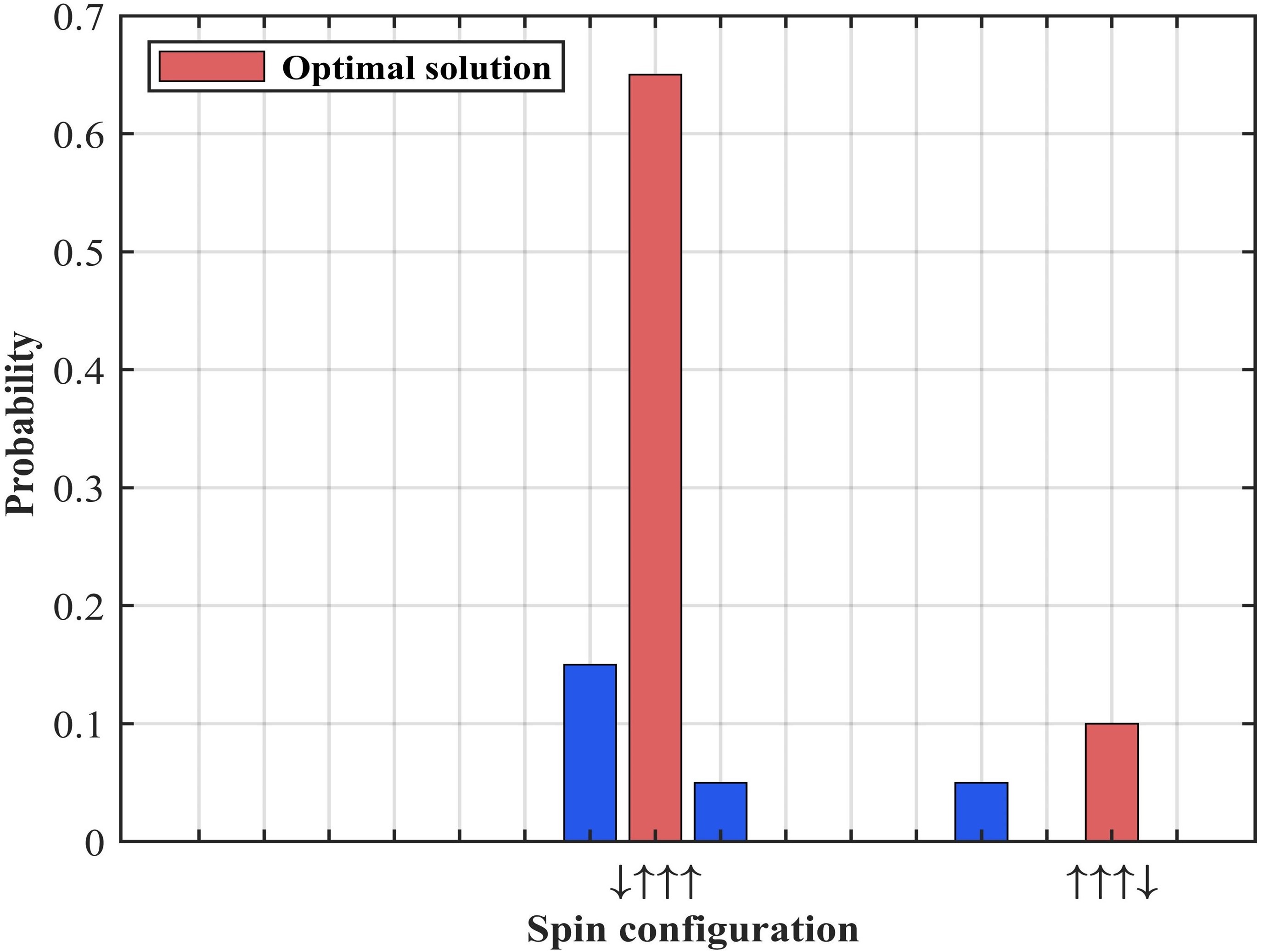}
\caption{\textbf{The 4-vertex cover problem.} The probability distribution of spin configuration against a 4-vertex cover problem and the final result converges to $|\downarrow\uparrow\uparrow\uparrow>$.}
\label{fig:vertexcover}
\end{figure}

\begin{figure}
\centering
\includegraphics[width=\columnwidth]{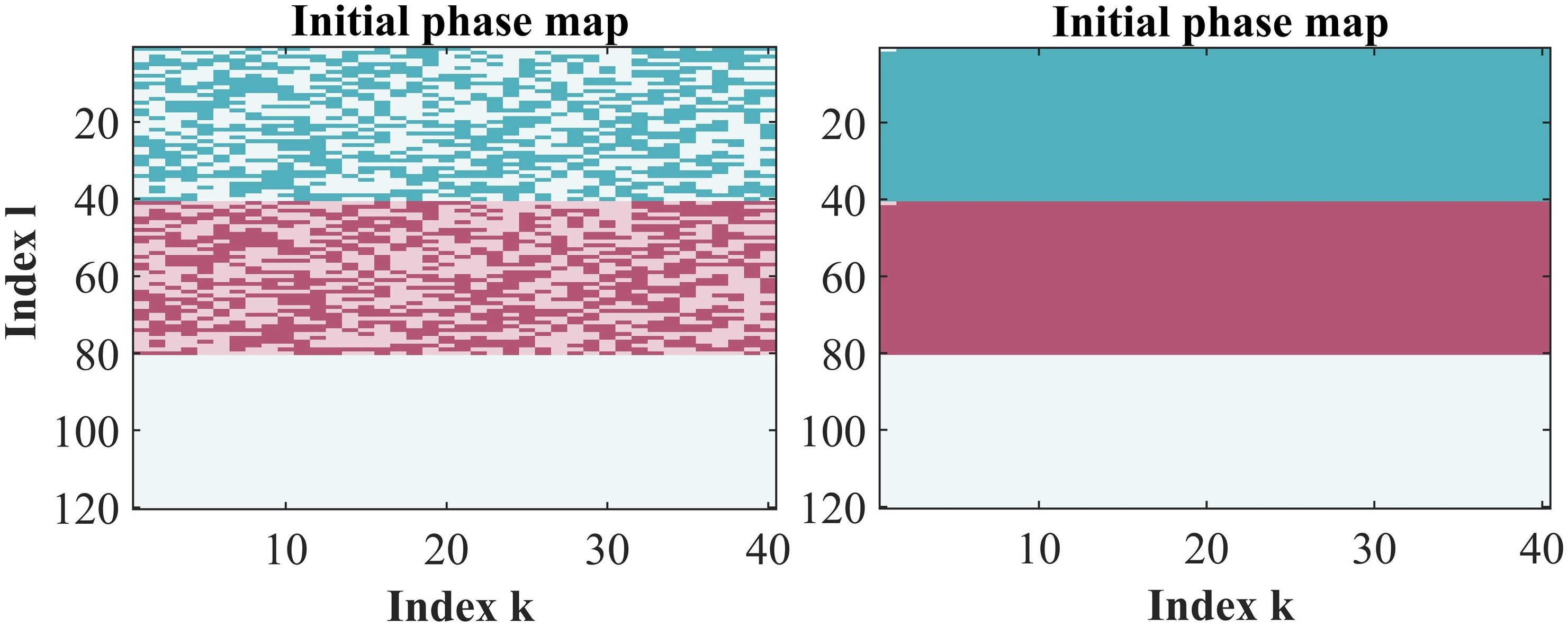}
\caption{\textbf{Lage-scale vertex cover problem.} A 1600-vertex cover problem is simulated. The left and right phase map is the initial and final phase solution  of our Ising machine.  
}
\label{fig:vertexcoverphasemap}
\end{figure}

We construct a vertex cover problem that scale is 4 and $A=B=4$ for demonstration. 
In experiment, spatial phase is directly modulated  for a  beam from laser source after expanding and shaping by a rectangular aperture. 
Relation matrix is set to be $diag(j,j,...,j)$. 
A fixed phase of $\pi$ is placed on the third section of SLM in order to obtain the last part of \eqref{Ising Hamiltonian}. 
During iterations, the probability of each spin configuration is shown in Fig. \ref{fig:vertexcover}. 
Because of a complete graph, the optimal solution is obvious as only one $|\downarrow>$ for any spin configurations. Fig. \ref{fig:vertexcover} demonstrates that our Q-SIM can complete the task with a high probability. 
A lager-scale vertex cover problem is simulated as shown in Fig. \ref{fig:vertexcoverphasemap}. 
In order to verify the solution, a complete graph is used for evaluation and Fig. \ref{fig:vertexcoverphasemap} shows the Q-SIM is able to solve a 1600-vertex cover problem with high probability. 

\section{Conclusion} 
To conclude, we propose a novel quadrature photonic spatial Ising machine to bridge over a chasm between NP hard problems and photonic Ising accelerator. 
The proposed architecture is able to 
configure arbitrary negative spin interaction through tuning the light amplitude and spatial phase based relation matrix, while preserving significant property of scalability.
Moreover, our scheme can run more optimization problems by configuring external magnetic field in Ising model.
Max-cut problem solution with graph order of 100 and density from 0.5 to 1 is experimentally demonstrated after almost 100 iterations, which is much less than integrated photonic Ising machine in the same scale because of spatial multiplexing.
Max-cut problem  with  more  than  1600  nodes  and  system tolerance for light misalignment are also investigated through numerical simulations.
Moreover, vertex cover problem has been successfully implemented by modeling photonic Ising model with external magnetic field.
Compared to DOPO based coherent Ising machine, the proposed spatial Ising machine has the scalability advantage but requires profound innovations of high-speed spatial phase modulators and detectors in order to reduce the iteration time.   
Our work pave a bright way for problem solution by large-scale photonic spatial Ising machine.

\section{Acknowledgement}
This work is supported by National Key Research and Development Program of China (2019YFB1802903).

The authors would like to thank Prof. Chihao Zhang for valuable discussions about the theory and implementation of Ising model.

\bibliography{Manuscript}

\end{document}